\documentclass[oneside,onecolumn]{article}

\usepackage{blindtext} 
\usepackage{graphicx}

\usepackage[sc]{mathpazo} 
\usepackage[T1]{fontenc} 
\usepackage{microtype} 

\usepackage[english]{babel} 

\usepackage[hmarginratio=1:1,top=32mm,columnsep=20pt]{geometry} 
\usepackage[hang, small,labelfont=bf,up,textfont=it,up]{caption} 
\usepackage{booktabs} 

\usepackage{lettrine} 

\usepackage{enumitem} 
\setlist[itemize]{noitemsep} 

\usepackage{abstract} 

\usepackage{titlesec} 
\renewcommand\thesection{\Roman{section}} 
\renewcommand\thesubsection{\roman{subsection}} 
\titleformat{\section}[block]{\large\scshape\centering}{\thesection.}{1em}{} 
\titleformat{\subsection}[block]{\large}{\thesubsection.}{1em}{} 

\usepackage{fancyhdr} 
\usepackage{titling} 

\usepackage{multirow}

\usepackage{hyperref} 
\usepackage{breakurl}

\pretitle{\begin{center}\Huge\bfseries} 
\posttitle{\end{center}} 
\title{Can the center lane really have advantages for swimmers?} 
\author{%
\textsc{Miwa Shirai} \\[1ex] 
\normalsize Meijo University \\ 
\normalsize \href{mailto:160441083@ccalumni.meijo-u.ac.jp}{160441083@ccalumni.meijo-u.ac.jp} 
\and 
\textsc{Eiji Konaka}\thanks{Corresponding author}\\[1ex] 
\normalsize Meijo University \\ 
\normalsize \href{mailto:konaka@meijo-u.ac.jp}{konaka@meijo-u.ac.jp} 
}
\date{\today} 
\renewcommand{\maketitlehookd}{%
\begin{abstract}
\noindent 
In swimming competitions, the center lane is considered to be the best lane. 
The main objective of this study is to discuss the validity of the assertion, ``good swimmers should be placed in the center lanes because these lanes has advantage.''
This study extracts the (positive) lane bias in the center lanes from many records utilizing statistical hypothesis tests. 
The data is collected from nationwide competitions of Japan over 10 years.
The data set contains more than 5900 swimmers and 56000 records.
No significant evidence was obtained on the  lane advantage of the center lanes.
\end{abstract}
}

\begin{document}

\maketitle


\section{Introduction}

Is there any ``lane advantage'' in swimming competitions?
The waves caused by the swimmers in both ends of the pool can reflect off the pool walls, and the reflected waves cause fluid resistance, which can worsen the performance, of other swimmers.
Therefore, the center lane is considered to be the best lane. 

In indoor swimming competitions, the assignment of lanes shall be by placing the fastest swimmer or team in the center lane in the pool.
The swimmer having the next fastest time is to be placed on his left, then alternating
the others to the right and left according to the submitted times \cite{FINASwimmingRules20172021}.

Swimmers cannot ignore fluid resistance caused by turbulence.
In addition to the reflection from the walls, reflection from the bottom of the pool
, and wave generated behind swimmers \cite{Yuan2019} may disturb a fair competition environment.

The modern swimming pool for the competition is designed for speed, with constant depth and overflow gutters to absorb turbulence generated by the swimmers \cite{colwin2002breakthrough}.
Most elite-level competitions are usually held with a pool with 10 lanes.
Usually, a maximum of  eight swimmers swim, where the lanes on both sides (lanes 0 and 10) are kept empty. These empty lanes also contribute to absorb the turbulence. 
Lane ropes with anti-wave functions have been designed and installed in modern pools \cite{4406822}. The anti-wave ropes prevent the waves generated by a swimmer from being transmitted to the next lane.

F\'{e}d\'{e}ration Internationale de Natation (FINA) designs swimming pools; in addition, FINA defines the upper and lower limitation of inflow and outflow to keep the water level constant. 
FINA also provides a practical test to check the amount of water movement \cite{FINAFacilitiesRules20172021}.
These improvements on facilities have been contributing in breaking the world records \cite{Nevill2008}.

Is the competition rule ``In pools using 10 lanes, the fastest swimmer shall be placed in lane
4 \cite{FINASwimmingRules20172021}'' necessary even today?
If these improvements on facilities described before reache a sufficiently high level, there is no reasonable basis to support this competition rule.
A coach of Olympic swimmers reported that recent technological improvement in designing pool and ropes almost diminished the advantage of the center lane \cite{Morgan}.

In contrast, the existence of lane advantage could not be rejected until now.
Measuring the effect of waves on the swimmers is challenging, even if a sufficient number of experiments to measure the performance of anti-wave lane ropes in the modern competition pool  are conducted.

Brammer et al. \cite{doi:10.1080/02640414.2016.1163402} and Cornett et al. \cite{Cornett2014} pointed out possible {\it lane bias}, i.e., improvement (or decline) in race record in specific lane and direction.
They claimed that a lane on one side, not center, in a temporary pool could have positive lane bias.
A detailed investigation based on fluid mechanics cannot be conducted for temporary pools because they are disassembled after a  competition.
Therefore, they used the statistical analysis method on race records after the competitions.

There are few academic investigations on lane advantage or lane bias in swimming.
The main objective of this study is to discuss the validity of the assertion, ``good swimmers should be placed in the center lanes because these lanes have an advantage.''
In particular, the existence of (positive) lane bias in the center lanes is discussed using many records and statistical hypothesis tests.

Swimming records in four elite-level nationwide competitions in Japan from 2007 to 2019 are collected.
The records consist of date, competition name, swimmer's name, style, lane number, heats or finals, and time.
The data set contains more than 5900 swimmers and 56000 records.
Suppose that the recorded time can be decomposed into the sum of the expected time for the swimmer, effect of heats or finals, and the lane effects.
A stepwise regression method is used to identify which predictor should be included in the model. 

This paper is organized as follows:
Section \ref{sec:data} describes the data set and the analysis method used in this study.
Section \ref{sec:result} shows the analysis results and discussions.
Section \ref{sec:conclusion} concludes this paper.


\section{Data and analysis method}
\label{sec:data}
In this section, the data set and analysis method used in this study are described in detail.

\subsection{Data set}
\label{sec:2-1}

The data source is \url{http://www.swim-record.com/}.
This site is officially approved by Japan Swimming Federation\footnote{\url{http://www.swim.or.jp/}}.
The following data set is collected:
\begin{itemize}
\item Years: 2007 to 2019
\item Competitions: Japan Swim, Japan Open, Inter-College Swimming Championship, National Sports Festival 
\item Style: Backstroke, Breaststroke, Butterfly, and Freestyle
\item Sex: Men and women
\item Distance: 50, 100, 200, and 400 m. 400 m is only for freestyle.
\item Record columns: Date, competition name, swimmer's name, style, lane number, heat/finals, and time.
\end{itemize}

Table \ref{tab:venue} lists all the venues of the competitions.
Twenty pools were used in these 52 competitions. Half of these (26/52) competitions were held at Tokyo Tatsumi International Swimming Center.
The National Sports Festival (in Japanese, {\it {Kokutai}}) is one of the biggest and most prestigious domestic multi-sport events in Japan.
The host prefecture was rotated throughout Japan. Therefore, every venue of this competition was different.

\begin{table}[h]
\caption{Venues}
\label{tab:venue}
\begin{center}
\begin{tabular}{p{6cm}cp{1cm}p{1cm}p{1cm}p{1.2cm}}\hline
Venue &  & Japan Swim & Japan Open & Inter-College & National Sports Festival \\ \hline
Tokyo Tatsumi International Swimming Center & 26 & 2008, 2010, 2012, 2014, 2015, 2016, 2018, 2019 & 2007, 2008, 2009, 2010, 2012, 2014, 2015, 2016, 2017, 2018, 2019 & 2007, 2008, 2010, 2012, 2016, 2019 & 2013 \\ \hline
ToBiO (Furuhashi Hironosin Memorial Hamamatsu Swimming Center) & 3 & 2009, 2011 &  & 2015 &  \\ \hline
Yokohama International Swimming Pool & 3 &  &  & 2011, 2014, 2018 &  \\ \hline
Daiei Probis Phoenix Pool & 2 & 2013 &  &  & 2009 \\ \hline
Chiba International General Swimming Cener & 2 & 2007 &  &  & 2010 \\ \hline
Kadoma Sports Center & 2 &  & 2011 & 2017 &  \\ \hline
Nagoya Civic General Gymnasium & 1 & 2017 &  &  &  \\ \hline
Sagamihara Green Pool & 1 &  & 2013 &  &  \\ \hline
Hiroshima Big Wave & 1 &  &  & 2013 &  \\ \hline
Aqua Dome Kumamoto & 1 &  &  & 2009 &  \\ \hline
Kasamatsu Sports Park & 1 &  &  &  & 2019 \\ \hline
Tsuruga City Sports Park & 1 &  &  &  & 2018 \\ \hline
Aqua Palette Matsuyama (Temporaly) & 1 &  &  &  & 2017 \\ \hline
Morioka City Pool & 1 &  &  &  & 2016 \\ \hline
Akibasan Pool & 1 &  &  &  & 2015 \\ \hline
Nagasaki Shimin Sogo Pool & 1 &  &  &  & 2014 \\ \hline
Nagaragawa Swimming Plaza & 1 &  &  &  & 2012 \\ \hline
Yamaguchi Kirara EXPO Memorial Park Swimming Pool & 1 &  &  &  & 2011 \\ \hline
Beppu City Aoyama Swimming Pool & 1 &  &  &  & 2008 \\ \hline
Akita Prefectural General Pool & 1 &  &  &  & 2007 \\ \hline
\end{tabular}
\end{center}
\end{table}
Table \ref{tab:datanum} lists the number of swimmers and records in the data set.
In this table, ``extracted'' and ``used'' mean the number of all extracted records and of used records in regression analysis mentioned later.
In the following regression analysis, we used records of swimmers with more than four records in one event (distance and style).
Outliers, often appeared in swim-offs, are excluded. The MATLAB function \verb|isoutlier|\footnote{\url{https://www.mathworks.com/help/matlab/ref/isoutlier.html}} is used to find outliers.
\begin{table}[htb]
\begin{center}
\caption{Size of data set}
\begin{tabular}{ccccccc} \hline
\multirow{2}{*}{Sex}
	&\multirow{2}{*}{Style} 
	&\multirow{2}{*}{Distance} 
	& \multicolumn{2}{c}{Swimmers} & \multicolumn{2}{c}{Records} \\ \cline{4-7}
	&&& extracted & used & extracted & used\\ \hline
	
Men&Backstroke&
		50&331 & 122 & 1494 & 824 \\
	&&100&714 & 258 & 3960 & 2616 \\
	&&200&617 & 255 & 3446 & 2357 \\
	&Breaststroke
	&50&376 & 133 & 1662 & 966 \\
	&&100&780 & 311 & 4383 & 3006 \\
	&&200&762 & 317 & 4279 & 2955 \\
	&Butterfly
	&50&367 & 135 & 1597 & 920 \\
	&&100&680 & 259 & 3827 & 2613 \\
	&&200&596 & 238 & 3295 & 2234 \\
&Freestyle&50 & 1132 & 306 & 4660 & 2745 \\
	&	&	100 & 1056 & 312 & 4680 & 2881 \\
	&	&200 & 756 & 253 & 3335 & 2029 \\
	&	&400 & 720 & 266 & 3904 & 2685 \\ \hline
Women		&Backstroke&
	50&319 & 132 & 1618 & 981 \\
	&&100&588 & 233 & 3761 & 2706 \\
	&&200&490 & 203 & 3091 & 2224 \\
	&Breaststroke
	&50&379 & 129 & 1672 & 993 \\
	&&100&624 & 241 & 3649 & 2588 \\
	&&200&557 & 223 & 3194 & 2250 \\
	&Butterfly
	&50&392 & 156 & 1703 & 978 \\
	&&100&614 & 253 & 3659 & 2586 \\
	&&200&513 & 216 & 3055 & 2195 \\
	&Freestyle
	&50 & 866 & 288 & 4343 & 2891 \\
	&	&100 & 1056 & 312 & 4680 & 2881 \\
	&	&200 & 593 & 201 & 2854 & 1822 \\
	&	&400 & 526 & 219 & 3169 & 2298 \\ \hline
	Total&&&
	16182	&5949&	84597&	56191\\ \hline

\end{tabular}
\label{tab:datanum}
\end{center}
\end{table}

\subsection{Analysis method}
\label{sec:2-2}


Let $T_{i,j,k}$ be the record time [s] of the swimmer $i$ swam in $j$-th ($j=0,\cdots, 9$) lane in heats or finals, which is denoted by $k=1,0$; 
$b_i$ means the expected time [s] for the swimmer $i$ in the corresponding competition.
The detailed definition of $b_i$ is provided later.

In the regression model, assume that the difference of the actual and the expected time, i.e, $T_{i,j,k}-b_i$, is explained by the linear combination of the heats/finals $k\in \{1,0 \}$, lane effects $l_{j,a}~(a=1,2)$, a constant $c$, and residual $\epsilon$.
\begin{equation}
T_{i,j,k}-b_i=Kh_k+\sum_{a=1}^4 A_a l_{j,a} +c+\epsilon, 
\end{equation}
where
\begin{equation}
\label{eqn:def_lja}
l_{j,a}=|j-4|^a.
\end{equation}
The lane effect terms $l_{j,a}$ is zero for $j=4$.
In this regression model, the lane effect becomes larger with respect to the distance from the fourth lane.
If $T_{i,j,k}-b_i$ is insensitive to the lane effect term, we can deduce that swimmers can not obtain an advantage from lane .

In this study, stepwise regression method is used to identify the elements that affects to the swim records. MATLAB function \verb|stepwiselm| \cite{slm-en} is used to determine that whether each predictor, i.e., $h_k,  l_{j,1}, \cdots , l_{j,4}, c$, should be included in the regression model or not.

In this study, default setting of \verb|stepwiselm| is used;
the criterion to add or remove terms $p$-value for an $F$-test of the change in the sum of squared error that results from adding or removing the term.

\begin{enumerate}
\item Start from the initial model, i.e., constant function.

\item Examine a set of available terms not in the model. 
If any of the terms have $p$-values less than an entrance threshold (that is, if it is unlikely a term would have a zero coefficient if added to the model), add the term with the smallest $p$-value and repeat this step; otherwise, go to step 3.

\item If any of the available terms in the model have $p$-values greater than an exit threshold (that is, the hypothesis of a zero coefficient cannot be rejected), remove the term with the largest p-value and return to step 2; otherwise, end the process.
\end{enumerate}


\subsection{Definition of expected time}
\label{sec:2-3}

Two candidates for the definition of the expected time $b_i$ for the swimmer $i$ are considered in this study.
\begin{itemize}
\item Mean time of the last 12 months.
\item Best time of the last 12 months.
\end{itemize}

For example, Rikako Ikee, who is the women's 50 and 100 meters national record holder as of 2021/3/5, participated in Japan Open at 2018/5/24. In the last 12 months before this competition, she had 5 records; 25.09, 24.33, 26.08, 24.21, and 24.75.
Therefore, $b_i=24.892$ or $24.33$ for each definition.
She recorded in this competition twice, 25.58 and 24.47, in the heat and final, respectively.

For each definition, basic statistics (mean, standard deviation, and skewness) of $T_{i,j,k}-b_i$ are listed in Table  \ref{tab:basicStats}. Because of the page limitation, this table lists only the statistics of freestyle.

\begin{table}[h]
\caption{Basic stats of $T_{i,j,k}-b_i$ for each definition of $b_i$ (freestyle)}
\label{tab:basicStats}
\begin{center}
\begin{tabular}{ccccc}\hline
&$b_i$& \multicolumn{3}{c}{Mean of the last 12 months}\\ \hline
Sex&Distance & Mean & Std & Skewness \\\hline
Men&50 & $-0.1073$  & 0.3213  & $-0.1404$  \\
&100 & $-0.1728$  & 0.5976  & $-0.1421$  \\
&200 & $-0.2494 $ & 1.2483  & 0.0351  \\
&400 & $-0.4731$  & 2.9066  & 0.0034  \\ \hline
Women&50 & $-0.0707$ & 0.3372 & $-0.0207$ \\
&100 & $-0.1339$ & 0.6446 & $-0.0353$ \\
&200 & $-0.2522$ & 1.5073 & $-0.0090$ \\
&400 & $-0.3404$ & 3.0204 & $-0.0024$ \\ \hline \hline
&$b_i$ & \multicolumn{3}{c}{Best of the last 12 months}\\ \hline
&Distance & Mean & Std & Skewness \\\hline
Men&50 & 0.0494  & 0.3728  & $-0.1113$ \\
&100 & 0.1343  & 0.6821  & $-0.0784$  \\
&200 & 0.3266  & 1.3278  & 0.0746  \\
&400 & 1.3208  & 3.3388  & 0.0205  \\\hline
Women& 50 & 0.1196 & 0.3908 & $-0.0464$ \\
&100 & 0.2047 & 0.7129 & $0.0224$ \\
&200 & 0.4718 & 1.6793 & $0.0204$ \\
&400 & 1.4879 & 3.5106 & $-0.0375$ \\ \hline

\end{tabular}
\end{center}
\end{table}

From this table, if $b_i$ is defined as the mean of the last 12 months, mean of $T_{i,j,k}-b_i$ is negative.
Therefore, it could be found that elite-level swimmers participating these nationwide competitions kept improving their records.
In other words, once the swimmers grow old and their record dropped,  they decided to retire from top-level competitions.

Unfortunately, the absolutes of the mean and the skewness of $T_{i,j,k}-b_i$ between both definitions has no consistency. Therefore, these basic statistics gives no evidence for which definitions should be adopted.
In this study, $b_i$ is defined as the best time of the last 12 months.

\subsection{Validation of regression model}

The regression model obtained by the stepwise regression method is validated by residual analysis.
In residual analysis, normality of residuals $\epsilon$ is tested.
If a distribution is normal, skewness and kurtosis for the samples from the distribution are $0$ and $3$, respectively.

Kolmogorov-Smirnov test is another test for normality \cite{10.2307/2280095}.
The one-sample Kolmogorov-Smirnov test is a nonparametric test of the null hypothesis that the sampled data is generated from the standard normal distribution.
MATLAB function {\tt{kstest}} \cite{kstest-en} performs this test.


\section{Analysis results and discussions}
\label{sec:result}
Tables from \ref{tab:resultBackstroke} to \ref{tab:resultFreestyle} lists the predictors added to the prediction model and the maximum lane effect. Lane effect is calculated for lanes numbered 1 to 8. 
The meaning of each column is as follows:
\begin{itemize}
\item Est.: Estimated coefficients
\item $p$-value: $p$-value for $t$-statistic for a test that the coefficient is zero  
\item L.E.[s]: Maximum lane effect in seconds
\item L.E.[m]: Maximum lane effect in length if the swimmer swims at Japanese national record time as of 2021/3/5.
\item Ratio to the JPN Record: Relative ratio to the Japanese national record
\end{itemize}

\begin{table*}[htb]
\begin{center}
\caption{Regression result: lane effect in backstroke}
\label{tab:resultBackstroke}
\begin{tabular}{c ccc c c c c} \hline
\multirow{2}{*}{Sex} &
\multirow{2}{*}{Distance} & & \multirow{2}{*}{Est.} &\multirow{2}{*}{$p-$value} &L.E.&L.E.& Ratio to the \\
& &  &  & & [sec] & [m] & JPN Record \\ \hline \hline
Men&50 & $h_k$ & $5.738\times10^{-2}$ & $4.45\times 10^{-4}$& - & - & - \\ \hline

\multirow{3}{*}{Men} &\multirow{3}{*}{100} 
	& $c$ & $1.106\times10^{-1}$ & $3.84\times 10^{-4}$& \multirow{3}{*}{$-0.068 $} &\multirow{3}{*}{$-0.130 $} & \multirow{3}{*}{$-1.30 \times 10^{-3}$} \\
	&& $l_{j,2}$ & $-4.246\times10^{-3}$ & $4.28\times 10^{-2}$ &  &  \\
	&& $h_k$ & $1.138\times10^{-1}$ &  $9.30\times 10^{-4}$ &  &  \\ \hline

Men&200 & $c$ & $5.417\times10^{-1}$ & $1.56\times 10^{-39}$& - & - & - \\ \hline

\multirow{2}{*}{Women} &\multirow{2}{*}{50} & 
	$c$ & $-7.318\times10^{-2}$ & $1.56\times 10^{-2}$ & \multirow{2}{*}{-} & \multirow{2}{*}{-} & \multirow{2}{*}{-} \\
	&& $h_k$ & $1.410\times10^{-1}$ &  $4.94\times 10^{-5}$&  &  \\ \hline

\multirow{2}{*}{Women} &\multirow{2}{*}{100} &
	 $c$ & $1.314\times10^{-1}$ & $3.61\times 10^{-5}$ & \multirow{2}{*}{-} & \multirow{2}{*}{-} & \multirow{2}{*}{-} \\
&& $h_k$ & $1.940\times10^{-1}$ &  $2.80\times 10^{-7}$&  &  \\ \hline

\multirow{3}{*}{Women} &\multirow{3}{*}{200} &
	$c$ & $2.571\times10^{-1}$ & $3.20\times 10^{-3}$ & \multirow{3}{*}{0.261} & \multirow{3}{*}{0.410} & \multirow{3}{*}{$2.05\times 10^{-3}$} \\
&& $l_{j,2}$ & $1.630\times10^{-2}$ &  $7.93\times 10^{-3}$&  &  \\ 
&& $h_k$ & $4.215\times10^{-1}$ &  $1.42\times 10^{-5}$&  &  \\ \hline

\end{tabular}
\end{center}
\end{table*}

\begin{table*}[htb]
\begin{center}
\caption{Regression result: lane effect in breasttroke}
\label{tab:resultBreaststroke}
\begin{tabular}{c ccc c c c c} \hline
\multirow{2}{*}{Sex} &
\multirow{2}{*}{Distance} & & \multirow{2}{*}{Est.} &\multirow{2}{*}{$p-$value} &L.E.&L.E.& Ratio to the \\
& &  &  & & [sec] & [m] & JPN Record \\ \hline \hline
Men&50 & $l_{j,3}$ &
	 $1.122\times10^{-3}$ & $1.11\times 10^{-3}$& $0.072$ & $0.133$ & $2.67\times 10^{-3}$ \\ \hline

Men&100 & $h_{k}$ &
	 $2.228\times10^{-1}$ & $1.41\times 10^{-32}$& - & - & - \\ \hline

\multirow{2}{*}{Men} &\multirow{2}{*}{200} &
	 $l_{j,4}$ & $5.980\times10^{-4}$ & $6.61\times 10^{-4}$ & \multirow{2}{*}{$0.153$} & \multirow{2}{*}{$0.242$} & \multirow{2}{*}{$1.21\times 10^{-3}$} \\
&& $h_k$ & $4.514\times10^{-1}$ &  $4.00\times 10^{-19}$&  &  \\ \hline

Women&50 & $l_{j,2}$ &
	 $6.438\times10^{-3}$ & $3.08\times 10^{-4}$& $0.103$ & $0.168$ & $3.36\times 10^{-3}$ \\ \hline

\multirow{4}{*}{Women} &\multirow{4}{*}{100} &
	 $c$ & $3.159\times10^{-1}$ & $1.73\times 10^{-6}$ & \multirow{4}{*}{$-0.247$} & \multirow{4}{*}{$-0.374$} & \multirow{4}{*}{$-3.74\times 10^{-3}$} \\
&& $l_{j,1}$ & $-1.466\times10^{-1}$ &  $2.74\times 10^{-3}$&  &  \\ 
&& $l_{j,2}$ & $2.148\times10^{-2}$ &  $2.56\times 10^{-2}$&  &  \\ 
&& $h_k$ & $3.004\times10^{-1}$ &  $4.30\times 10^{-9}$&  &  \\ \hline

Women&200 & $h_{k}$ &
	 $9.820\times10^{-1}$ & $4.19\times 10^{-45}$& - & - & - \\ \hline

\end{tabular}\end{center}
\end{table*}

\begin{table*}[htb]
\begin{center}
\caption{Regression result: lane effect in butterfly}
\label{tab:resultButterfly}
\begin{tabular}{c ccc c c c c} \hline
\multirow{2}{*}{Sex} &
\multirow{2}{*}{Distance} & & \multirow{2}{*}{Est.} &\multirow{2}{*}{$p-$value} &L.E.&L.E.& Ratio to the \\
& &  &  & & [sec] & [m] & JPN Record \\ \hline \hline

\multirow{3}{*}{Men} &\multirow{3}{*}{50} &
	$l_{j,1}$ & $-4.247\times10^{-2}$ & $4.96\times 10^{-5}$ & \multirow{3}{*}{$-0.094$} & 		\multirow{3}{*}{$-0.202$} & \multirow{3}{*}{$-4.05\times 10^{-3}$} \\
&& $l_{j,4}$ & $4.097\times10^{-2}$ &  $7.42\times 10^{-6}$&  &  \\ 
&& $h_k$ & $1.407\times10^{-1}$ &  $1.26\times 10^{-9}$&  &  \\ \hline

\multirow{4}{*}{Men} &\multirow{4}{*}{100} &
	 $c$ & $7.320\times10^{-2}$ & $1.39\times 10^{-2}$ & \multirow{4}{*}{$-0.057$} & \multirow{4}{*}{$-0.115$} & \multirow{4}{*}{$-1.12\times 10^{-3}$} \\
&& $l_{j,3}$ & $-5.834\times10^{-3}$ &  $3.84\times 10^{-3}$&  &  \\ 
&& $l_{j,4}$ & $1.242\times10^{-3}$ &  $9.15\times 10^{-4}$&  &  \\ 
&& $h_k$ & $2.193\times10^{-1}$ &  $3.17\times 10^{-12}$&  &  \\ \hline

\multirow{2}{*}{Men} &\multirow{2}{*}{200} &
	 $c$ & $2.194\times10^{-1}$ & $5.76\times 10^{-4}$ & \multirow{2}{*}{-} & \multirow{2}{*}{-} & \multirow{2}{*}{-} \\
&& $h_k$ & $3.810\times10^{-1}$ &  $6.15\times 10^{-7}$&  &  \\ \hline

\multirow{2}{*}{Women} &\multirow{2}{*}{50} &
	 $c$ & $-1.318\times10^{-1}$ & $5.23\times 10^{-6}$ & \multirow{2}{*}{-} & \multirow{2}{*}{-} & \multirow{2}{*}{-} \\
&& $h_k$ & $2.302\times10^{-1}$ &  $3.86\times 10^{-12}$&  &  \\ \hline

Women&100 & $h_{k}$ &
	 $3.655\times10^{-1}$ & $1.10\times 10^{-77}$& - & - & - \\ \hline

\multirow{3}{*}{Women} &\multirow{3}{*}{200} &
	$c$ & $4.244\times10^{-1}$ & $1.10\times 10^{-4}$ & \multirow{3}{*}{$-0.358$} & 		\multirow{3}{*}{$-0.574$} & \multirow{3}{*}{$-2.87\times 10^{-3}$} \\
&& $l_{j,1}$ & $-8.943\times10^{-2}$ &  $8.59\times 10^{-3}$&  &  \\ 
&& $h_k$ & $7.321\times10^{-1}$ &  $2.28\times 10^{-12}$&  &  \\ \hline

\end{tabular}\end{center}
\end{table*}

\begin{table*}[htb]
\begin{center}
\caption{Regression result: lane effect in freestyle}
\label{tab:resultFreestyle}
\begin{tabular}{c ccc c c c c} \hline
\multirow{2}{*}{Sex} &
\multirow{2}{*}{Distance} & & \multirow{2}{*}{Est.} &\multirow{2}{*}{$p-$value} &L.E.&L.E.& Ratio to the \\
& &  &  & & [sec] & [m] & JPN Record \\ \hline \hline
\multirow{3}{*}{Men} &\multirow{3}{*}{50} 
	& $l_{j,1}$ & $-2.064\times10^{-2}$ & $1.52\times 10^{-4}$& \multirow{3}{*}{$-0.051$} &\multirow{3}{*}{$-0.117$} & \multirow{3}{*}{$-2.34\times 10^{-3}$} \\
	&& $l_{j,4}$ & $1.382\times10^{-4}$ & $4.12\times 10^{-3}$ &  &  \\
	&& $h_k$ & $1.117\times10^{-1}$ &  $1.18\times 10^{-19}$ &  &  \\ \hline

Men&100 & $h_k$ & $1.806\times10^{-1}$ & $6.85\times 10^{-33}$& - & - & - \\ \hline

\multirow{2}{*}{Men} &\multirow{2}{*}{200} & $c$ & $1.179\times10^{-1}$ & $3.44\times 10^{-2}$ & \multirow{2}{*}{-} & \multirow{2}{*}{-} & \multirow{2}{*}{-} \\
&& $h_k$ & $2.888\times10^{-1}$ &  $3.60\times 10^{-6}$&  &  \\ \hline

\multirow{4}{*}{Men} &\multirow{4}{*}{400} & $l_{j,1}$ & $-1.673\times 10^{-1}$ & $1.46\times 10^{-2}$ &
	 \multirow{4}{*}{$-0.366$} &  \multirow{4}{*}{$-0.653$} &  \multirow{4}{*}{$-1.63\times 10^{-3}$}\\
& & $l_{j,4}$ &$1.683\times 10^{-3}$ & $3.60 \times 10^{-4}$\\
& &$c$ & $1.006\times 10^0$ & $9.37\times 10^{-10}$ &  & &  \\
&& $h_k$ & $6.822\times 10^{-1}$ &  $3.60\times 10^{-6}$ &  &  \\ \hline

Women&50 & $h_k$ & $1.725\times10^{-1}$ & $7.07\times 10^{-87}$& - & - & - \\ \hline
Women&100 & $h_k$ & $2.869\times10^{-1}$ & $1.24\times 10^{-71}$& - & - & - \\ \hline
Women&200 & $h_k$ & $6.658\times10^{-1}$ & $9.82\times 10^{-44}$& - & - & - \\ \hline

\multirow{2}{*}{Women} &\multirow{2}{*}{400} & $c$ & $6.337\times10^{-1}$ & $2.56\times 10^{-2}$ & \multirow{2}{*}{-} & \multirow{2}{*}{-} & \multirow{2}{*}{-} \\
&& $h_k$ & $1.203\times10^{0}$ &  $6.69\times 10^{-14}$&  &  \\ \hline

\end{tabular}
\end{center}
\end{table*}

Table \ref{tab:residualError} summarizes skewness, kurtosis, and $p$-value of Kolmogorov-Smirnov test for residuals.

\begin{table}[h]
\caption{Residual error analysis}
\label{tab:residualError}
\begin{center}
\begin{tabular}{ccrrrr}\hline
Sex&Style&Distance & Skewness & Kurtosis & $p$-value for KS test\\\hline
Men &Backstroke
	&50 & $-0.0611$ & $3.5195$ & $0.2349$ \\
	&&100 & $-0.0069$ & $3.1019$ & $0.2867$ \\
	&&200 & $0.1459$ & $3.1463$ & $0.0922$ \\
	&Breaststroke
	&50 & $-0.0810$ & $3.0511$ & $0.0702$ \\
	&&100 & $0.0135$ & $3.3278$ & $*0.0017$ \\
	&&200 & $-0.0020$ & $3.2171$ & $*0.0240$ \\
	&Butterfly
	&50 & $-0.1049$ & $3.1954$ & $0.0785$ \\
	&&100 & $-0.0589$ & $3.2185$ & $*0.0048$ \\
	&&200 & $0.0997$ & $3.1249$ & $*0.0126$ \\
 &Freestyle
	&50 & $-0.1340$  & 3.0661  & $*$0.0299\\
	&&100 & $-0.0951$  & 3.1313  &0.1527\\
	&&200 & 0.0354  & 3.2530  &$*$0.0251\\
	&&400 & 0.0014  & 3.1790  &0.1256
\\\hline
Women &Backstroke
	&50 & $-0.0853$ & $3.1944$ & $0.1511$ \\
	&&100 & $0.0571$ & $3.1720$ & $0.2297$ \\
	&&200 & $0.0449$ & $3.1926$ & $0.0327$ \\
	&Breaststroke
	&50 & $-0.0666$ & $2.9339$ & $0.3995$ \\
	&&100 & $-0.1005$ & $3.0863$ & $0.0924$ \\
	&&200 & $-0.1328$ & $3.1163$ & $*0.0055$ \\
	&Butterfly
	&50 & $-0.1293$ & $2.9178$ & $0.2381$ \\
	&&100 & $-0.1172$ & $3.1995$ & $*0.0031$ \\
	&&200 & $0.0675$ & $3.0523$ & $0.0789$ \\
 &Freestyle
	&50 & $-0.0729$ & $2.9754$ & $0.6786$ \\
	&&100 & $-0.0141$ & $3.0758$ & $*0.0432$ \\
	&&200 & $-0.0151$ & $3.1787$ & $0.0910$ \\
	&&400 & $-0.0867$ & $3.1511$ & $0.0674$ \\
\hline
\end{tabular}

$*$: $p<0.05$
\end{center}
\end{table}

\subsection{Discussions}
In most of the events (23 from 26), $h_k$, which denotes heats or finals, is included as a predictor in prediction model with small $p$-value (smaller than $10^{-3}$).


The predictors that express the lane effects, i.e., $l_{j,1}, \cdots , l_{j,4}$, are included in only 11 from 26 events. Even in these events, $p$-values of these lane effect terms are larger than those of $h_k$.
This means that $l_{j,1}, \cdots , l_{j,4}$ have much less importance than $h_k$ to explain $T_{i,j,k}-b_i$.
The sign of the largest lane effect is not consistent between events.
Only four events shows positive lane advantage for the center lanes. 
The estimated largest effect is at most $0.5\%$ with respect to the Japanese national records, even if the lane effect exists.

\begin{figure}[h]
\centering
\begin{tabular}{cc}
\includegraphics[width=0.45\columnwidth]{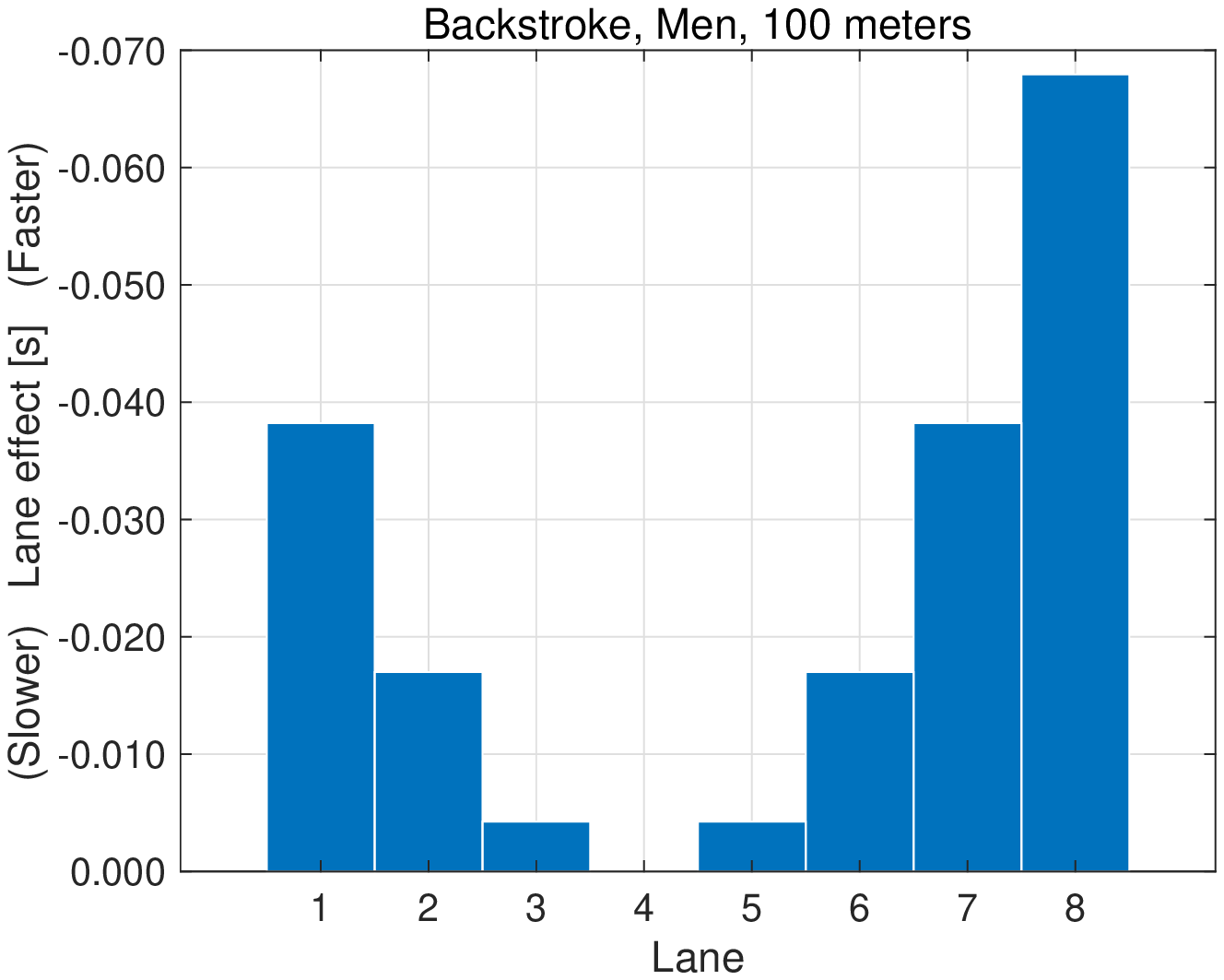}
&
\includegraphics[width=0.45\columnwidth]{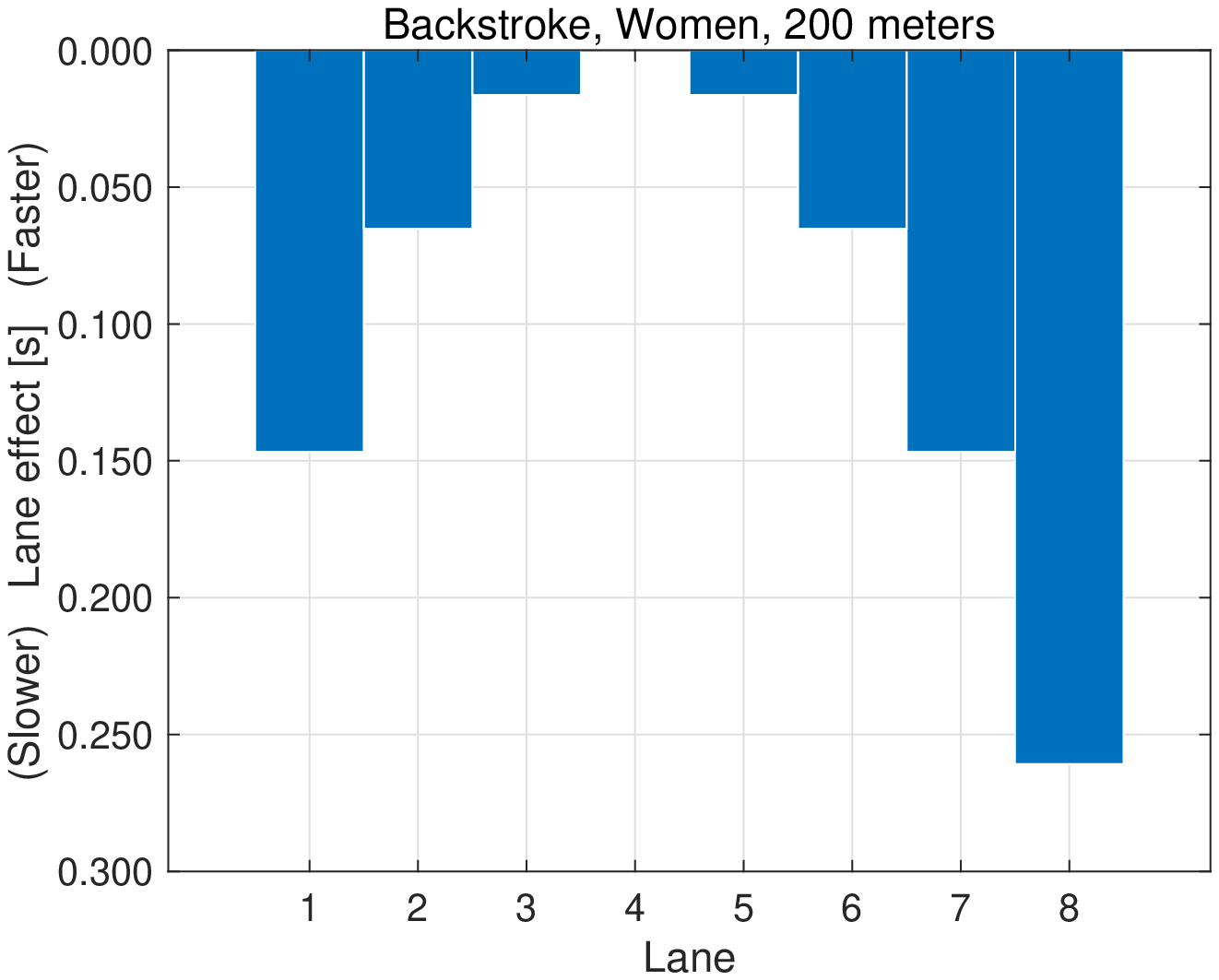}\\
(a)&(b)
\end{tabular}
\caption{Lane effect, Backstroke (Men's 100 and Women's 200 meters)}
\label{fig:laneEffectBackstroke}
\end{figure}

\begin{figure}[h]
\centering
\begin{tabular}{cc}
\includegraphics[width=0.45\columnwidth]{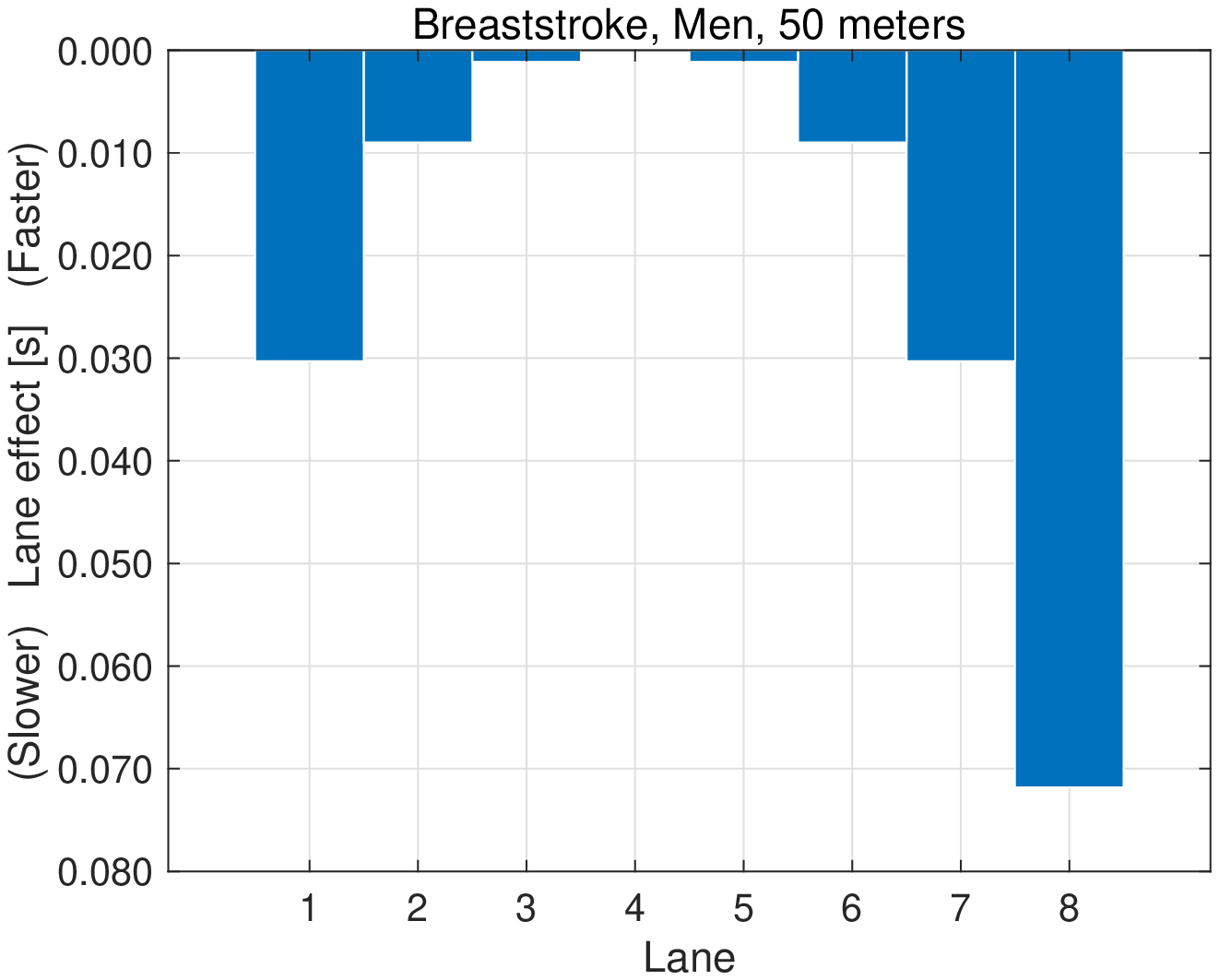}
&
\includegraphics[width=0.45\columnwidth]{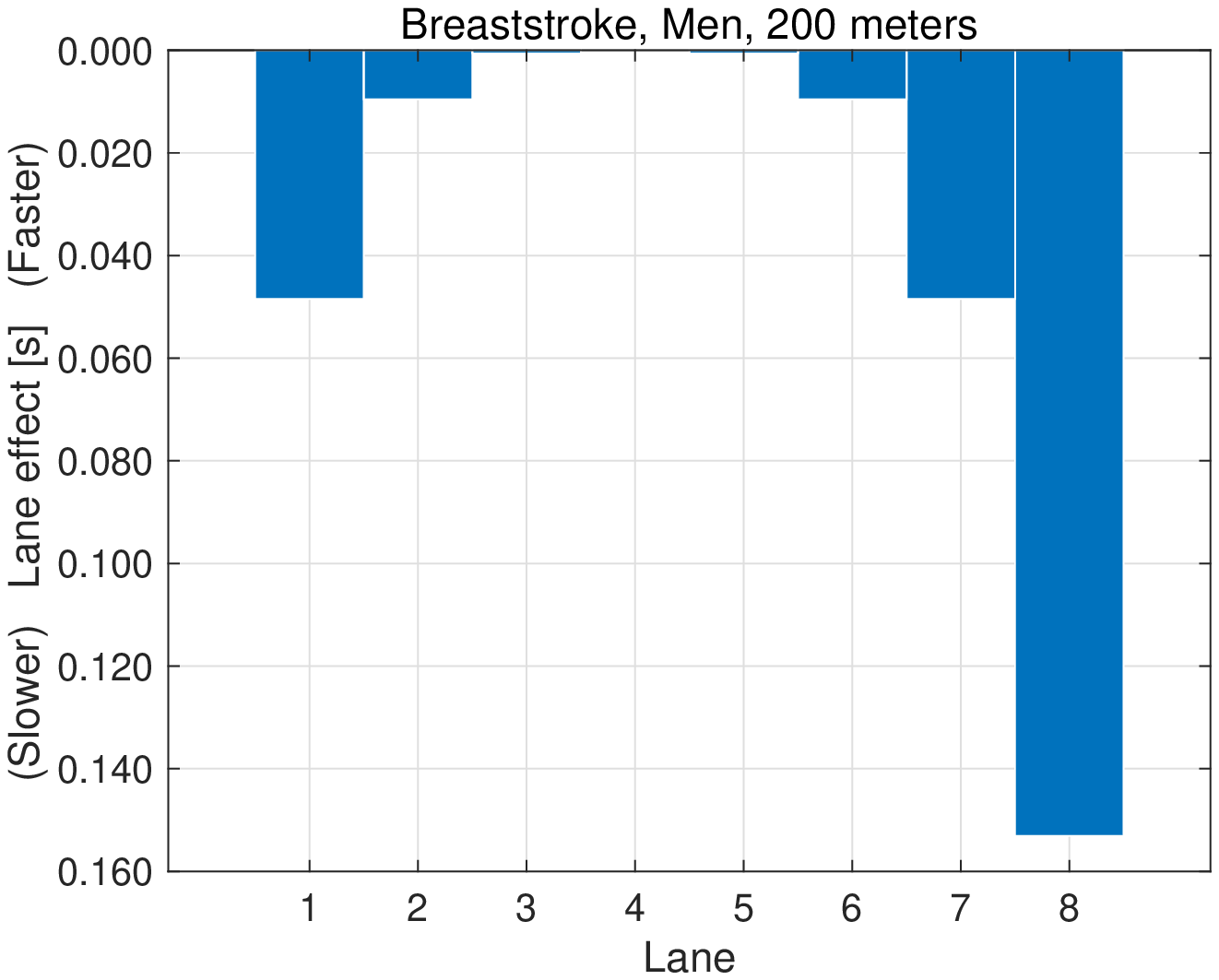}\\
(a)&(b)\\ \\
\includegraphics[width=0.45\columnwidth]{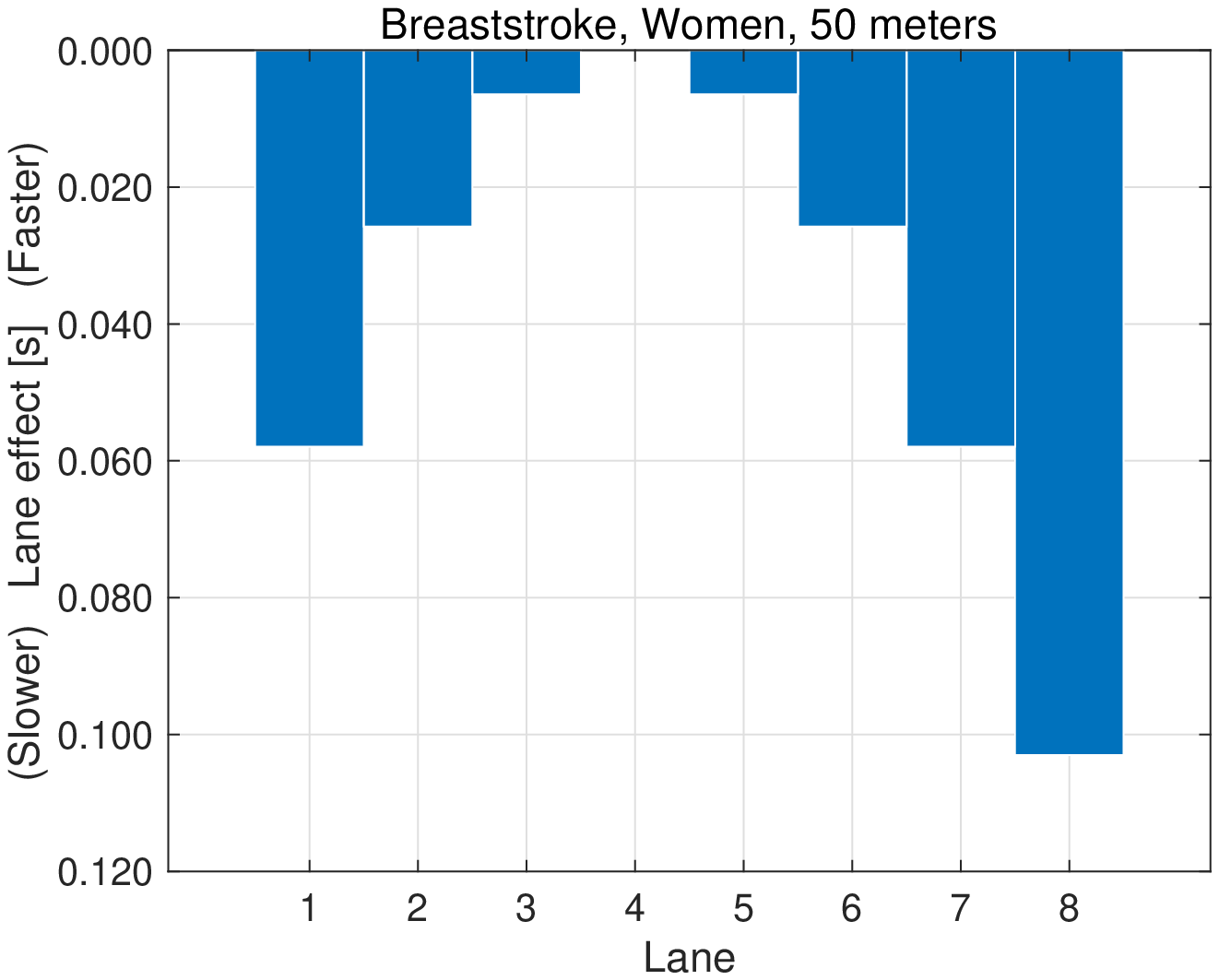}
&
\includegraphics[width=0.45\columnwidth]{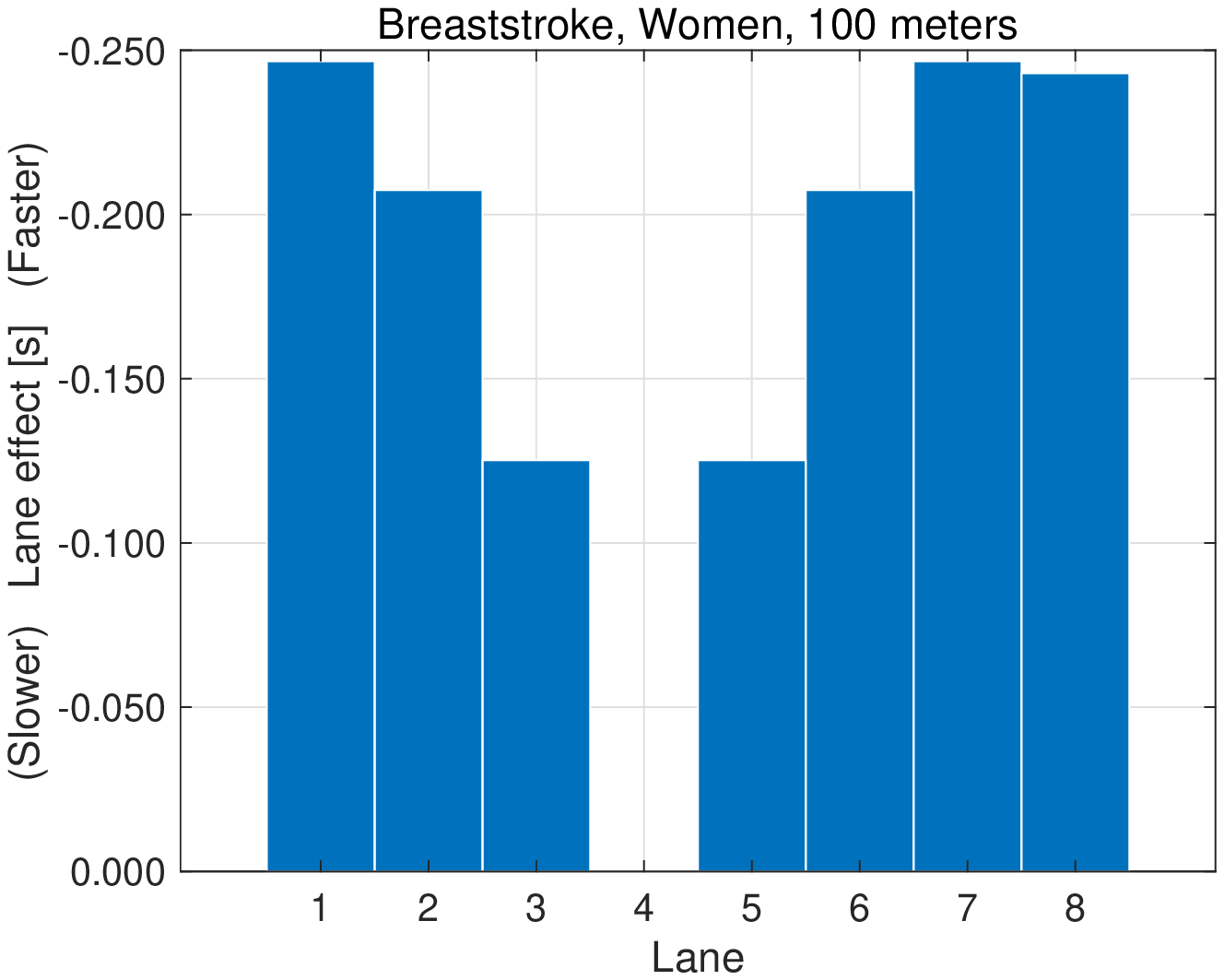}\\
(c)&(d)\\
\end{tabular}
\caption{Lane effect, Breaststroke (Men's 50 and 200, and Women's 50  and 100 meters)}
\label{fig:laneEffectBreaststroke}
\end{figure}

\begin{figure}[h]
\centering
\begin{tabular}{cc}
\includegraphics[width=0.45\columnwidth]{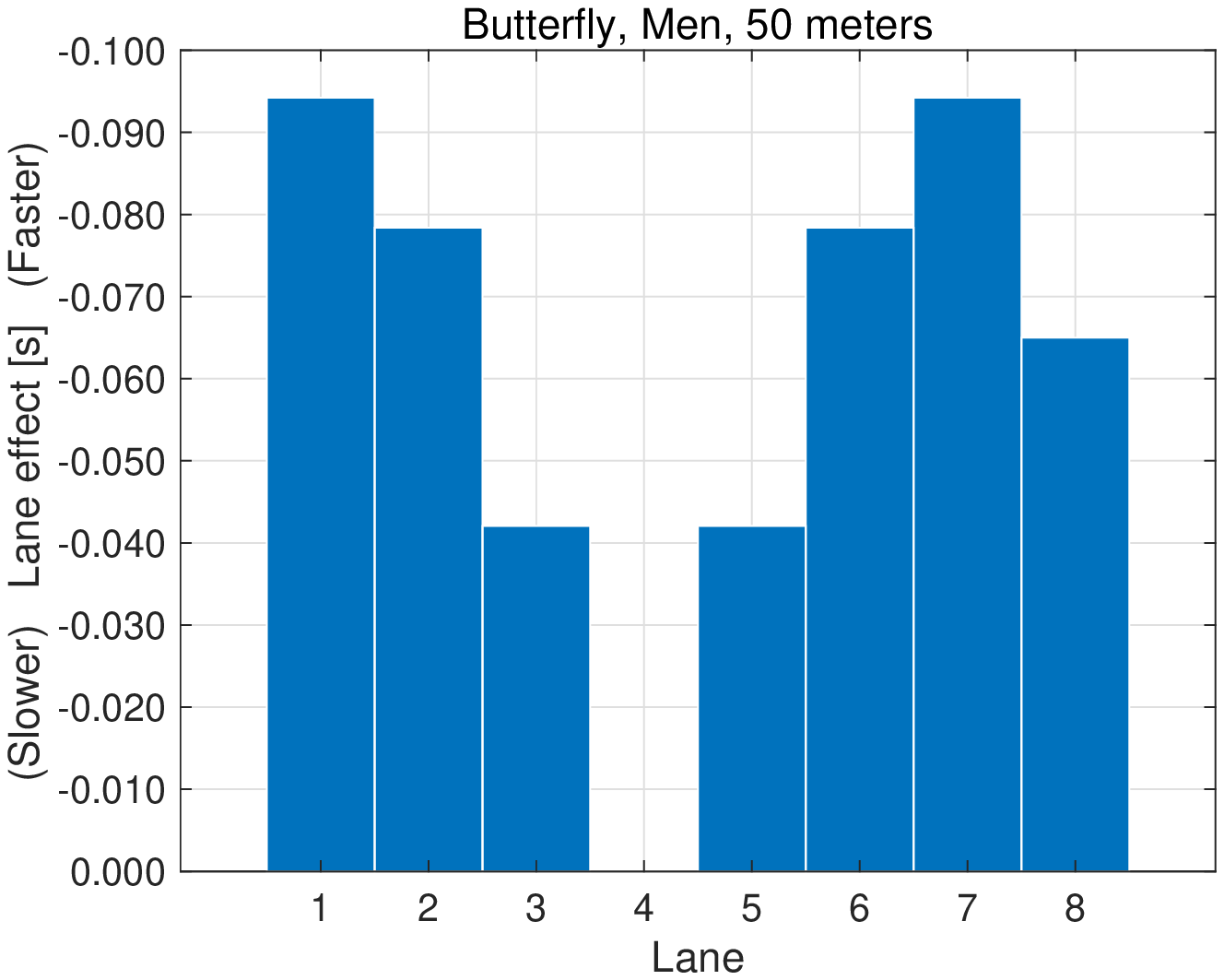}
&
\includegraphics[width=0.45\columnwidth]{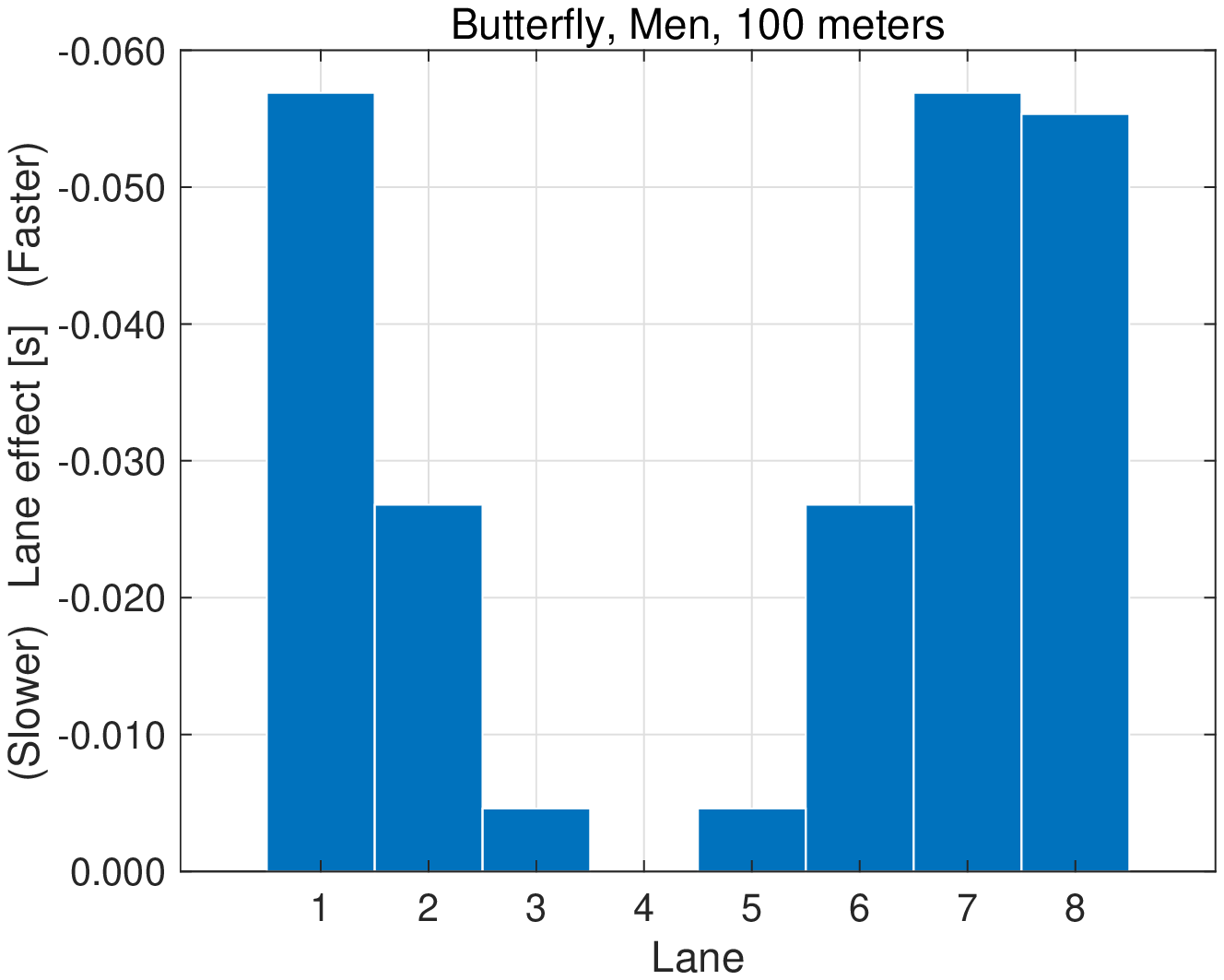}\\
(a)&(b)\\ \\
\includegraphics[width=0.45\columnwidth]{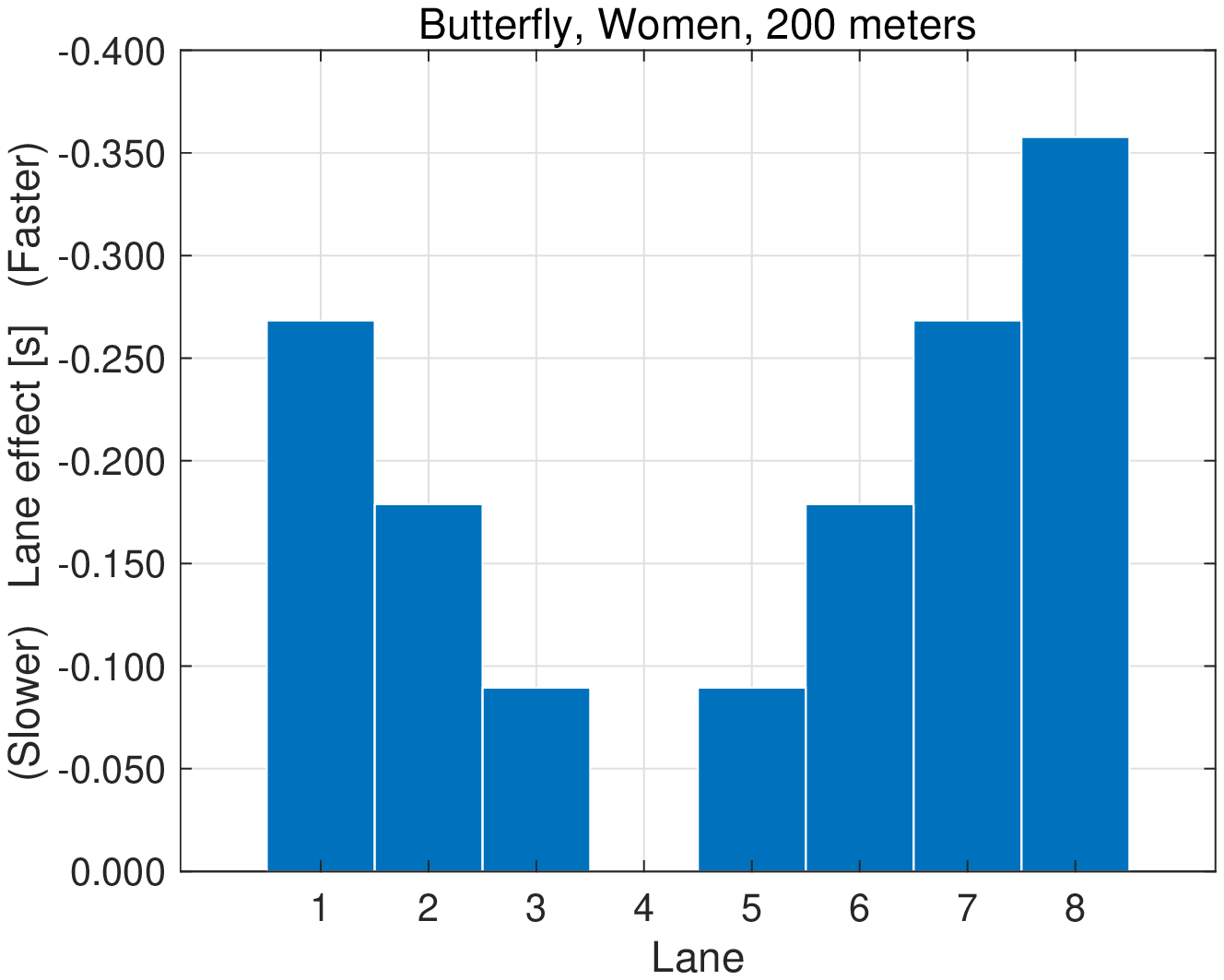}\\
(c)
\end{tabular}
\caption{Lane effect, Butterfly (Men's 50 and 100, and Women's 200)}
\label{fig:laneEffectButterfly}
\end{figure}

\begin{figure}[h]
\centering
\begin{tabular}{cc}
\includegraphics[width=0.45\columnwidth]{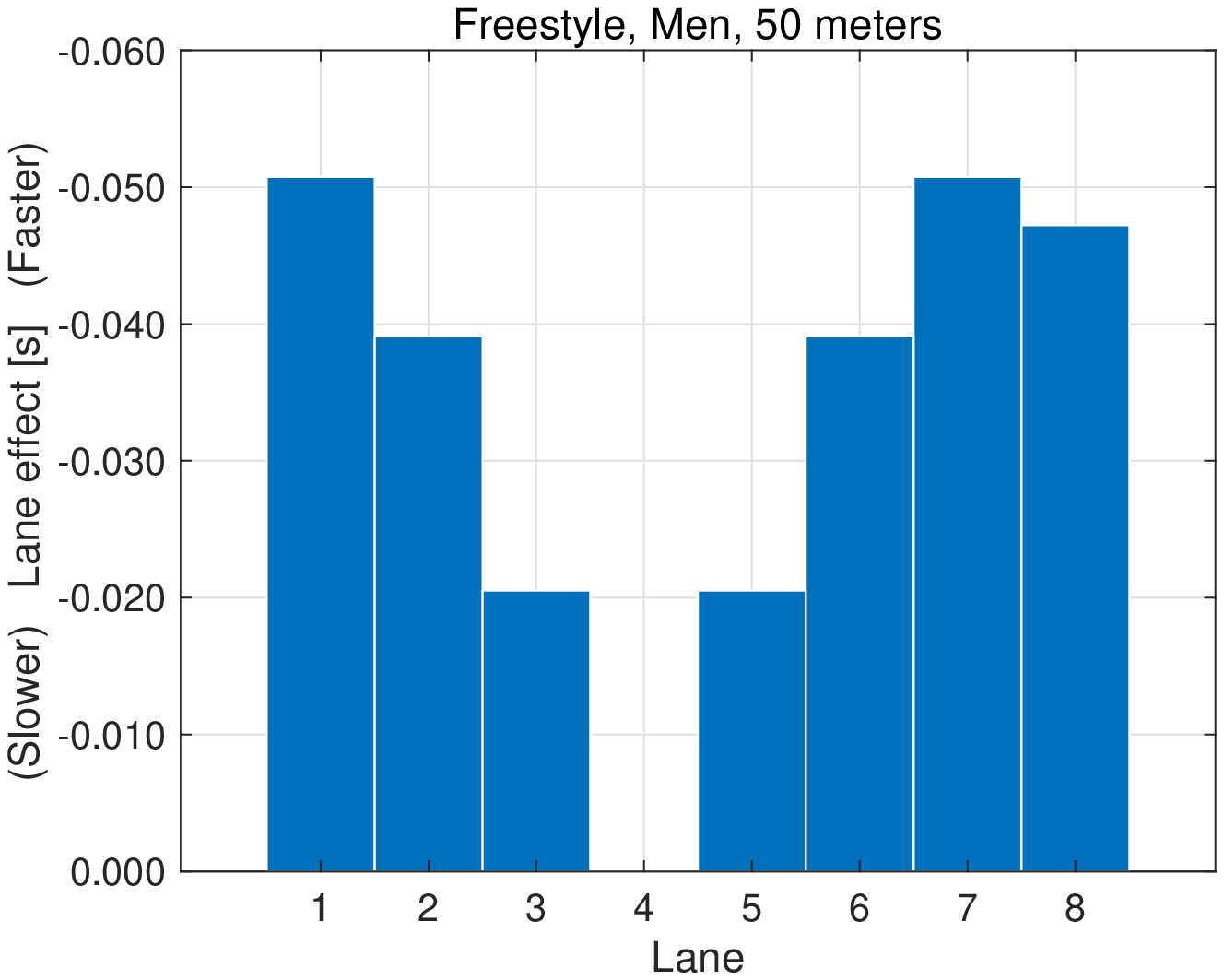}
&
\includegraphics[width=0.45\columnwidth]{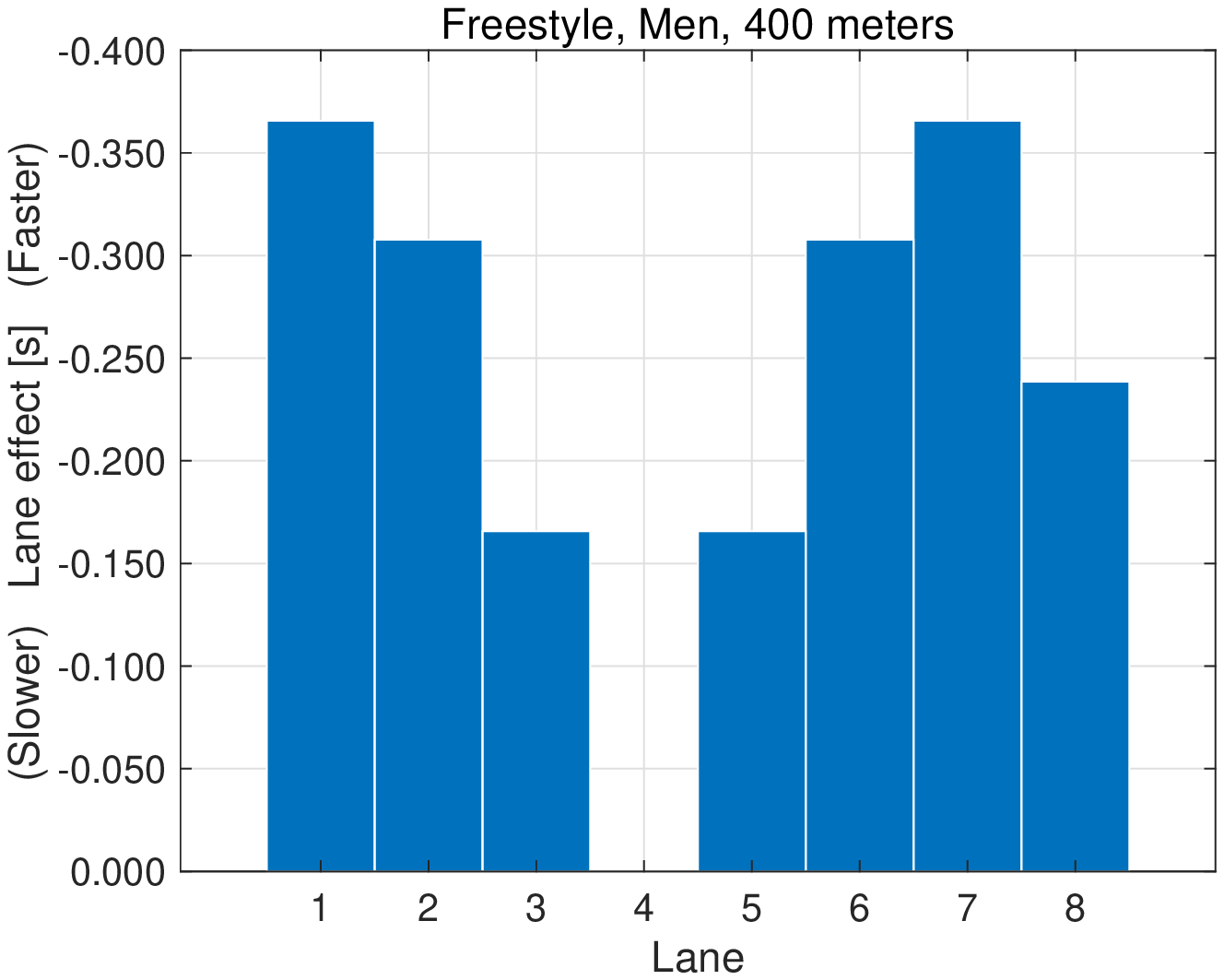}\\
(a)&(b)
\end{tabular}
\caption{Lane effect, Freestyle (Men's 50 and 400 meters)}
\label{fig:laneEffectFreestyle}
\end{figure}

Figures from \ref{fig:laneEffectBackstroke} to \ref{fig:laneEffectFreestyle} illustrate the estimated lane effect.
These figures show that the center lanes have disadvantage in butterfly and freestyle.


Table \ref{tab:residualError} shows that all absolute values of skewness are less than $0.5$, therefore all distributions can be considered as symmetric.
In addition, all kurtosis values are near $3.0$, that of the ideal normal distribution.
In only nine events, $p$-values of KS test are less than $0.05$.
Based on these results, it cannot be concluded that these residual distributions do not always follow normal distribution.  

As a result, this statistical analysis cannot find significant evidence that supports the assertion, ``good swimmers should be placed in the center lanes because these lanes have advantage.''
At the same time, this study cannot reject this assertion.
The current swimmer assignment rule is neither reasonable or unreasonable.

\section{Conclusion}
\label{sec:conclusion}
In this study, discuss the validity of the assertion, ``good swimmers should be placed in the center lanes because these lanes have advantage.''
Especially, the existence of (positive) lane bias in the center lane is discussed using large amount of records and statistical hypothesis test.
The data set contatins more than 5900 swimmers and 56000 records are used.
As a result, this statistical analysis cannot find significant evidence that supports the existence of the lane advantage in the center lane.




\begin{thebibliography}{10}

\bibitem{FINASwimmingRules20172021}
{FINA}.
\newblock {FINA SWIMMING RULES}, September 2017.

\bibitem{Yuan2019}
Zhi-Ming Yuan, Mingxin Li, Chun-Yan Ji, Liang Li, Laibing Jia, and Atilla
  Incecik.
\newblock Steady hydrodynamic interaction between human swimmers.
\newblock {\em Journal of The Royal Society Interface}, 16, January 2019.

\bibitem{colwin2002breakthrough}
Cecil Colwin.
\newblock {\em Breakthrough swimming}, pages 223--225.
\newblock Human kinetics, 2002.

\bibitem{4406822}
Nadim-Pierre Rizk.
\newblock Wave-damping properties of swimming lines.
\newblock Master's thesis, LUND UNIVERSITY, 2013.

\bibitem{FINAFacilitiesRules20172021}
{FINA}.
\newblock {FINA FACILITIES RULES}, September 2017.

\bibitem{Nevill2008}
Alan Nevill, Greg Whyte, R~Holder, and Michael Peyrebrune.
\newblock Are there limits to swimming world records?
\newblock {\em International Journal of Sports Medicine}, 28:1012--7, 01 2008.

\bibitem{Morgan}
FIona Morgan.
\newblock Are there lane advantages in athletics, swimming and track cycling?
\newblock BBC newsbeat
  (\url{http://www.bbc.co.uk/newsbeat/article/37083059/are-there-lane-advantages-in-athletics-swimming-and-track-cycling}),
  August 2016.
\newblock accessed 2020/9/8.

\bibitem{doi:10.1080/02640414.2016.1163402}
Christopher Brammer, Andrew Cornett, and Joel Stager.
\newblock Lane bias in elite-level swimming competition.
\newblock {\em Journal of Sports Sciences}, 35(3):283--289, 2017.
\newblock PMID: 27019224.

\bibitem{Cornett2014}
Andrew Cornett, Chris Brammer, and Joel Stager.
\newblock Current controversy: Analysis of the 2013 fina world swimming
  championships.
\newblock {\em Medicine and science in sports and exercise}, 47:649--654, 07
  2015.

\bibitem{slm-en}
MathWorks.
\newblock stepwiselm. {Perform} stepwise regression ({MATLAB} online
  documentation).
\newblock \url{https://www.mathworks.com/help/stats/stepwiselm.html}.
\newblock accessed 2020/9/8.

\bibitem{10.2307/2280095}
Frank~J. Massey.
\newblock The kolmogorov-smirnov test for goodness of fit.
\newblock {\em Journal of the American Statistical Association},
  46(253):68--78, 1951.

\bibitem{kstest-en}
Mathworks.
\newblock kstest. {One}-sample {Kolmogorov-Smirnov} test ({MATLAB} online
  documentation).
\newblock \url{https://www.mathworks.com/help/stats/kstest.html}.
\newblock accessed 2020/9/8.

\end{thebibliography}


\end{document}